# Large FCC colloidal crystals under microgravity

Qian Lei[1], Boris Khusid[1], Lou Kondic[2], Paul M. Chaikin[3], Andrew D. Hollingsworth[3†], Alton J. Reich[4], William V. Meyer[5]

[1]Otto H. York Dept. of Chemical and Materials Engineering, New Jersey Institute of Technology
[2]Dept. of Mathematical Sciences, New Jersey Institute of Technology
[3]Dept. of Physics, New York University
[4]Streamline Automation LLC, Huntsville, AL
[5] USRA at NASA Glenn Research Center, Cleveland, OH

[†]Corresponding author: Andrew Hollingsworth, New York University; Dept. of Physics, 726 Broadway, NY, NY 10003; 212-998-8428; ah95@nyu.edu

## Abstract

Here we report the results of microgravity experiments performed on the International Space Station (ISS) to study crystallized, metastable colloidal liquids in the region of the hard sphere phase diagram that has been found to be glassy on Earth. Using confocal microscopy, we observed the self-assembly of high density, three-dimensional colloidal crystals from micron-sized hard spheres suspended within a fluid medium. The largest face-centered cubic (FCC) phase measured 27 x 1.2 x 0.15 mm. From a practical aspect, the fact that a single, topological defect-free FCC colloidal crystal can be grown in microgravity and the crystal returned from orbit suggests new routes for manufacturing colloidal devices, particularly optical elements in space.

## Introduction

The goal for the Advanced Colloids Experiment with Temperature Control-11 (ACE-T11) ISS experiment was to study hard-sphere colloidal dispersions at particle volume fractions above $\phi \sim 0.56$. At these concentrations, samples generally remain disordered or 'glassy' on Earth because of gravitational sedimentation and particle jamming (Hunter and Weeks, 2012). The experiments were carried out using micron-sized poly(methyl methacrylate) (PMMA) spheres dispersed in a refractive index matching oil, and the effective hard sphere volume fraction was set by the predicted disorder-order transition.

### *Crystallization rate and structure*

Our Physics of Hard Spheres Experiment (PHaSE) microgravity experiment confirmed that FCC is the stable equilibrium structure, exhibiting slow growth rates during crystal annealing (Cheng *et al.*, 2002). Recent computer simulations showed that the interfacial free energy of the HCP solid may be slightly higher than that of the FCC phase, resulting in a greater likelihood that FCC will nucleate from the (metastable) liquid compared with HCP (Sanchez-Burgos *et al.*, 2021). The ACE-T11 microgravity experiments, as well as earlier studies by our group (Zhu *et al.*, 1997), were designed to investigate the nature of phase transitions in colloids in the absence of gravity. The results reported here were achieved at the particle level using an established optical imaging technique for increasing optical resolution and contrast by means of using a spatial pinhole array to block out-of-focus light during image formation.

In addition to theoretical interest, we envision taking advantage of the self-assembly of colloidal materials—the spontaneous organization of microscopic particles into ordered structures—to produce materials ideally with targeted properties. The implementation of this strategy allows colloidal particles to function as building blocks of new materials in myriad technologies including materials engineering, electronics, and optics, as well as life science, chemical and pharmaceutical industries.

## Materials and Methods

The colloidal particles were synthesized using established dispersion polymerization methods (Antl *et al.*, 1986; Pathmamanoharan *et al.*, 1997). The dispersant required to prevent particle aggregation was prepared according to Elsesser and Hollingsworth (2010). This polymeric stabilizer is poly(12-hydroxystearic acid)-*graft*-poly(methyl methacrylate) or PHS-*g*-PMMA, a comb-graft copolymer dissolved in an acetates and toluene mixture at about 40 weight percent solute. The ACE-T11 samples featured a rhodamine-fluorescein hybrid dye called julolidine rhodol (Sauers *et al.*, 1987; Dickinson *et al.*, 2010), which proved to be superior to commercially-available rhodamine isothiocyanate (RITC). We measured the excitation and emission spectra, which are presented in Figure 1, along with a vertical line denoting the frequency-doubled Nd:YAG (532 nm) laser system in the Light Microscopy Module (LMM). The dye was esterified with an acrylate monomer in order to co-polymerize it with methyl methacrylate and methacrylic acid. An SEM image of the PMMA particles is presented in Figure 2.

Next, the particles were transferred to a refractive index fluid, typically a mixture of decahydronaphthalene (45% w/w) and tetrahydronaphthaline (55% w/w). The decalin/tetralin mixture matched the (equilibrated) particle and fluid refractive indices, $n_D$ = 1.5097 at a wavelength of 589 nm. In this process, the particle and fluid masses were weighed carefully to produce the desired effective volume fraction. A heat shock step was required to equilibrate the 'good' solvent (tetralin) with the PMMA (Kodger *et al.*, 2017). Following this step, one or two test samples were prepared at just below and above the freezing volume fraction to observe colloidal crystals in the coexistence phase (Pusey and van Megen, 1986). Finally, the effective volume fraction of the samples, $\phi_{eff}$, was determined by constructing the phase diagram using the sedimentation method and calibrating with respect to the known freezing point, *i.e.* setting 0.494 as the lowest volume fraction which first shows crystallization (Phan *et al.*, 1996). This step identifies the PHS layer thickness, typically 10 to 15 nm, and volumetric swelling ratio of 1.29.

The particle suspensions were loaded into individual sample cells by capillary action. Exterior surfaces were wiped off and each tube was sealed using indium metal (Belser, 1954). A drop of epoxy was then applied to the ends of each cell. The flat capillaries (Electron Microscopy Science P/N 63825-05; Corning 7740 borosilicate glass capillary, 0.20 x 2.0 x 50 mm) were mounted on a copper thermal bridge equipped with thermistors to control a temperature gradient across the length of the capillary. Sample loading and mounting into two sample modules was conducted at NYU and at ZIN Technologies' facilities. Module SN 2017 is shown in Figure 3 along with the corresponding sample volume fractions contained within capillaries C1–C3. Module SN R002 was filled four months later for another ACE-T experiment that will be reported separately.

The sample modules were delivered to the ISS on December 4, 2019 by way of SpaceX-19. Confocal microscopy was conducted during September–October 2020 and again in January–February 2021. The LMM flight unit features a modified Leica DMRXA microscope configured with a Yokogawa CSU-X1 confocal scanner and a Q-Imaging Retiga 1300 camera. A 100x oil immersion objective was used to image the colloidal samples. The LMM was operated remotely from the NASA GRC Payload Operations Center (GIPOC) without the crew participation. We used ImageJ and MATLAB software for image processing and analysis of data collected in the ISS experiments.

**Results**

Three concentrated samples were prepared for the ACE-T11 experiment: $\phi_{eff} = 0.52$ (C1, coexistence), 0.55 (C2, crystal), 0.59 (C3, glassy). An average, unswollen particle diameter of 1.25 μm (4.3% PDI) was determined from scanning electron microscope (SEM) image analysis of more than 500 particles. The particle suspensions were homogenized in 1-g just prior to launch aboard the Space-X Falcon 9 rocket. At $\phi_{eff} = 0.59$, the nucleation and growth of colloidal crystals are greatly hindered in a gravitational field and we expect that particles would remain in an amorphous state. Indeed, experimental evidence suggests that a hard-sphere glass transition, if it exists, would occur in the range $\phi \sim 0.56$ (random loose packing) to $\phi \sim 0.64$ (random close packing) (Hunter and Weeks, 2012).

A glassy phase was not observed in C3, which gradually crystallized upon exposure to the microgravity environment on the ISS. In this sample cell, a face-centered cubic (FCC) phase, 27 x 1.2 x 0.15 mm single colloidal crystal was grown from solid, spherical particles as indicated by the confocal microscopy images presented in Figures 4 and 5. The confocal images shown in Figures 5a and 5b are actually a superposition of three 109.4 x 80.4 micron scans collected in the *x*-*y* plane, in which about 50% of each full image has been removed periodically due to uneven illumination. The three images corresponding to different *x*-coordinate locations were merged in three channel (red/green/blue) format to form a composite image panel, five of which are shown in Figure 5b. The image manipulation demonstrates structural coherence as indicated by the consistency of positions and orientation of six bright spots on the Fourier space pattern obtained by a FFT algorithm. The hexagonal symmetry corresponds to periodic elements, *i.e.*, the exposed (111) facet of a space-filling FCC colloidal crystal.

For the ACE-T11 samples in C1 and C2, the particle sediment proved to be too viscous to manipulate the metal stir wires. In C2, small crystallites did form however, within a 5–10 micron layer near the top cover. We imposed a vertical temperature gradient intended to induce sample mixing resulting from thermophoresis. This gradient was produced by coupling the oil objective maintained at ambient temperature to a heated capillary creating a ~40°C temperature difference across the 0.2 mm capillary channel. However, no significant particle displacement was observed after several hours in either C1 or C2.

We note that the ACE-T11 experimental apparatus was designed to apply less intense temperature gradients across the length of the capillary cell to explore the effect of thermophoresis on crystallization of hard-sphere colloids. Remarkably, we could not detect a Soret effect—thermal diffusion induced by an imposed temperature gradient—as described, for example, by Piazza and Parola (2008).

The NJIT and NYU investigators worked in close collaboration with researchers from the NASA Glenn Research Center to record, process and analyze ~130 GB confocal microscopy images of colloids collected in the ISS experiments in 2020–2021.

**Discussion**

Our previous Space Shuttle experiments indicated that high volume fraction hard sphere particle suspensions might self-assemble into an ordered crystalline structure on ISS (Zhu *et al.* 1997). With this observation in mind, the ACE-T11 experiment was developed to explore the role of macroscopic and microscopic forces in structure formation. To mediate crystallization, the ACE-T11 study included heating to produce elevated sample temperatures, as well as a reversible temperature gradient to induce particle concentration gradients (Cheng *et al.*, 1999). This feature was intended to allow the disorder-to-order transition to be observed at the particle level—that is, a change in number density could have been followed if the particle concentration adjusted itself throughout the sample cell. Scanning across a capillary in the direction of increasing particle density, *i.e.*, towards the cold side, we expected to see the formation of crystallites in fluid (coexistence) and further along, a space-filling crystal. At the cold end of the capillary, a sufficiently high volume fraction would produce a glassy phase on Earth, whereas a crystal structure could possibly result in the microgravity environment.

Colloidal crystallization in the glassy phase is due to the absence of gravity that, on Earth, causes particle jamming, sedimentation, and convection. Concentrated and disordered monodisperse colloidal particles forming crystals in microgravity was observed previously, but at the macroscopic level (Zhu *et al.* 1997). In the Colloidal Disorder-Order Transition (CDOT) experiments, the brilliant sample iridescence was attributed to the diffraction and constructive interference of visible light waves according to Bragg's law. That is, interstitial voids between periodic arrays of spherical PMMA particles (508 or 518 nm diameter) acted as a natural diffraction grating for visible light waves. Thus, *physical color*—indicative of structure—occurred when particle crystallites formed. In contrast, ACE-T11 used confocal microscopy to observe crystallization of micron-sized spheres at the particle level.

Although we were not able to induce crystallization using a temperature gradient across C1 or C2, the colloid in capillary C3 formed a FCC crystal nearly the size of the capillary used to contain it. FCC crystalline regions were found in capillaries C1 and C2 only near the top surface of the capillary. Furthermore, the colloidal samples used in the ISS experiments were sufficiently concentrated to survive re-entry and remained crystalline upon return to Earth, although minor structural disturbance occurred during transport. Most notably during the ACE-T11 experiment, a large crystal, like nothing seen on Earth, was created on ISS. At this particle density, the sample would have remained as a disordered glass on Earth. Yet in microgravity, the colloid contained within capillary C3 exhibited a large 3-D crystal, made from micron-sized particles, without visible structural defects. These experimental results should help guide the fabrication of 3D photonic structures.

Since the particles comprising these large, ordered structures (crystals) are the size of the wavelength of visible light, they have the potential to control light without resorting to heat producing electronics which limits miniaturization. One exciting aspect of this work is the realization that in the future we may be able to send up large quantities of these particles—monodisperse collections with different diameters, shapes and properties based on the wavelength of light we wish to control with these crystals—in these types of solvents, which do not seem to be subject to Soret effects. Upon return to Earth, these concentrated samples may exhibit desirable large 3-D, defect-free structure. Ultimately, this experiment should enable scientists to map a pathway for fabrication of multiphase micro-structured heterogeneous items from polymers, metals and ceramics. The study also forms the technological foundation for the next generation of complex processes like high-resolution additive manufacturing (3-D printing).

## Conclusions

Results of the ACE-T11 experiments demonstrated that molecular dynamics simulations of equilibrium phase diagram are valid only when gravitational effects are not important. Measurement of the motion of individual particles in crystalline and non-crystalline regions of colloids thus provides a unique opportunity for testing and validating various assumptions underlying molecular dynamics simulations of the structure and dynamics of simple liquids.

Physical realizations of the hard sphere system are almost possible with colloidal spheres dispersed in solvents of the same index of refraction. Unfortunately, index matching (to diminish the attractive van der Waals interaction and multiple scattering) is usually incompatible with density matching. Gravity tends to cause inevitable settling, preventing a uniform density at equilibrium. Moreover, the imposition of control fields such as temperature gradients possibly leads to instabilities of a fluid phase.

Our ACE-T11 experiments show clear differences between the phase diagram of hard spheres, crystal structure and crystallite morphology for samples measured in 1-g and those measured in micro-g. These differences are of intrinsic interest on their own, *e.g.*, the facility of crystallization in micro-g of samples which remain in the glass phase in 1-g. The implication is that dynamical processes at volume fractions near close packing are extremely sensitive to gravity.

The main effects of gravity can be seen from the Boltzmann factor governing the particle density for dilute systems as a function of height, $exp\left(\frac{-\Delta mgh}{k_B T}\right)$, where $\Delta m$ is the buoyant mass. In our samples, a gravitational height, $h = \frac{k_B T}{\Delta mg}$, as high as 30 microns implies, to lowest order, a factor of 20 density difference from the bottom to the top of a 100 micron-thick section. To reduce this difference to an acceptable level (~0.002 variation in concentration), we must reduce the average acceleration of gravity to $10^{-4}$ g. Osmotic pressure effects increase the effectiveness of thermal energy and change the real requirement to ~$10^{-3}$ g (Schmidt *et al.* 2004).

The contrast in structures formed in model particle suspensions under microgravity in the ACE-T11 experiment and standard gravity on Earth revealed the salient features of the influence of a temperature gradient and gravity on non-equilibrium colloidal processes. Understanding these phenomena is essential for the development and operation of a wide range of terrestrial and space applications involving the (spontaneous or directed) self-assembly of mesoscopic materials. The particles should be dispersed in fluids which do not exhibit Soret effects (thermophoresis). The particle sizes, shapes and properties can be varied based on the wavelength of light we wish to control with colloidal photonic crystals.

Three dimensional photonic crystals have potential applications ranging from: (i) Zero-threshold microlasers with high modulation speed; (ii) Low-threshold optical switches and all optical transistors for optical telecommunication; (iii) High-speed optical computers; (iv) Microlasers operating near a photonic band edge that will exhibit ultra-fast modulation and switching speeds for application in high-speed data transfer and computing; (v) Telecommunications with light without the need for conversion to and from electrical signals; (vi) Multiple scattering of light in biological tissue providing a safe, inexpensive and non-insidious probe of brain, breast and skin tumors (Dissanayake and Wijewardena Gamalath, 2015; Johri *et al.*, 2007).

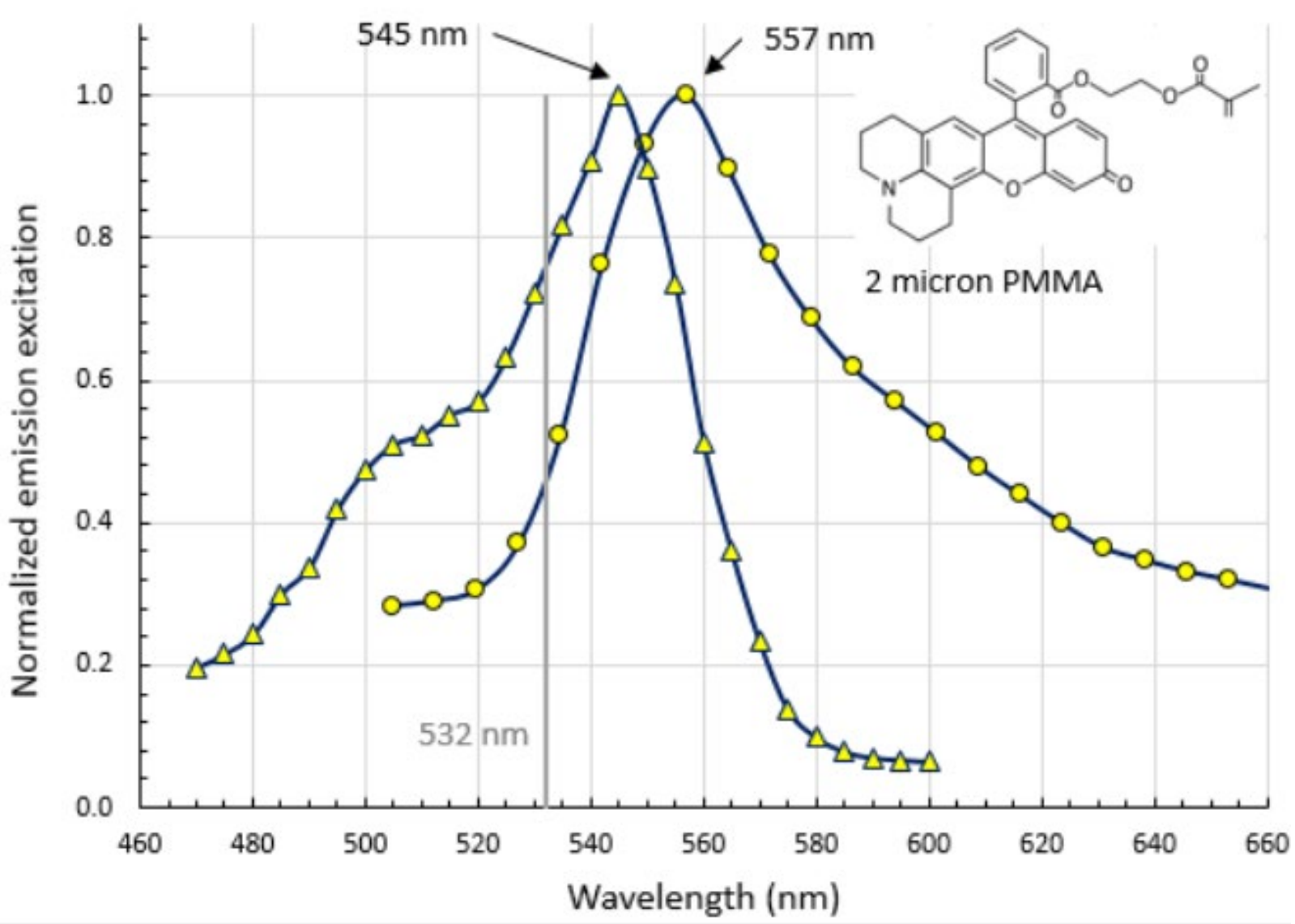

**Figure 1.** Excitation and emission spectra for julolidine rhodol-labeled (2 micron diameter) PMMA particles. Measurements were made using a Leica SP8 confocal laser scanning microscope.

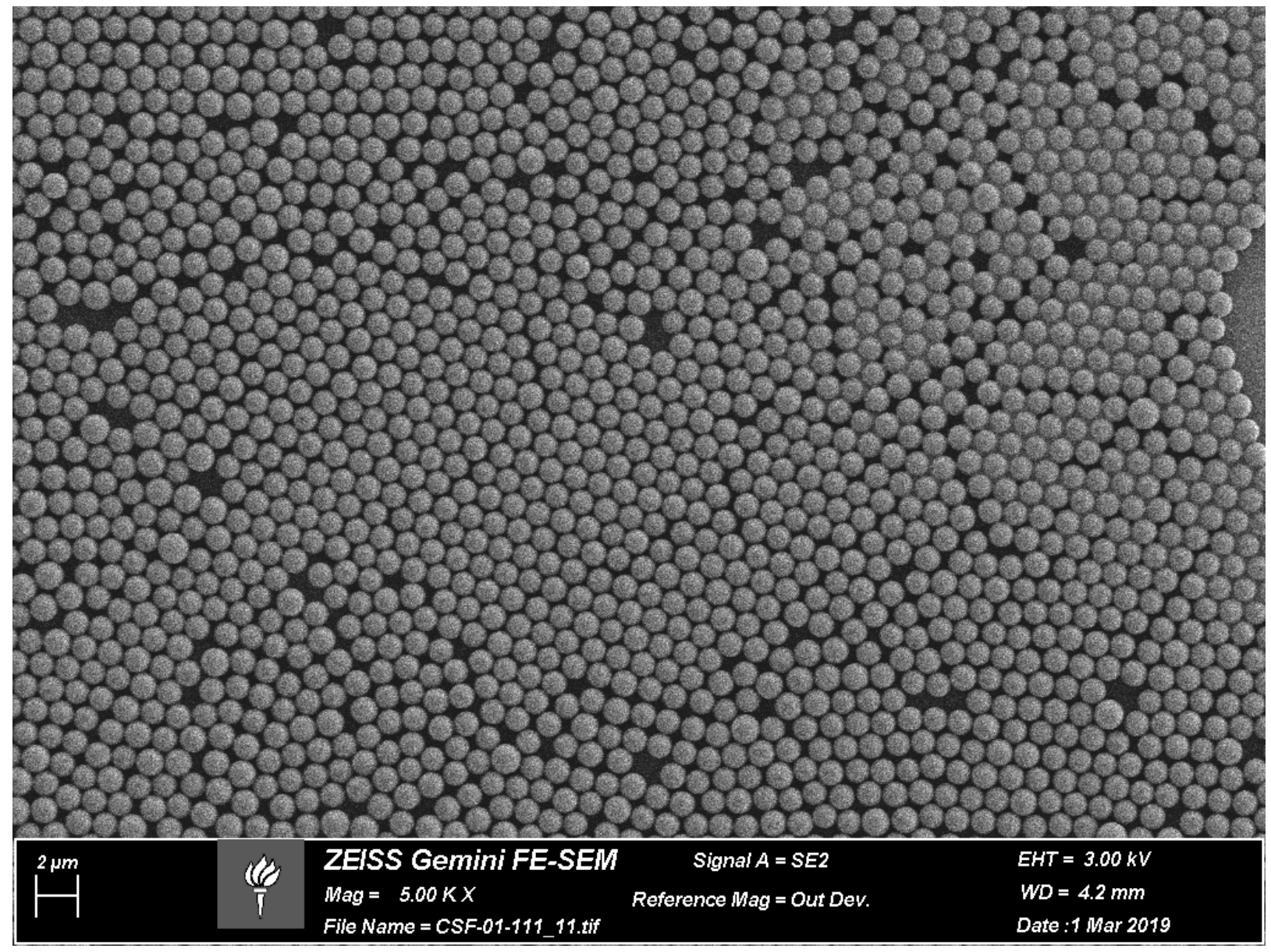

**Figure 2.** SEM image of the 1.25 µm average diameter PMMA spheres (4.3% PDI) using a MERLIN (Carl Zeiss) field emission scanning electron microscope.

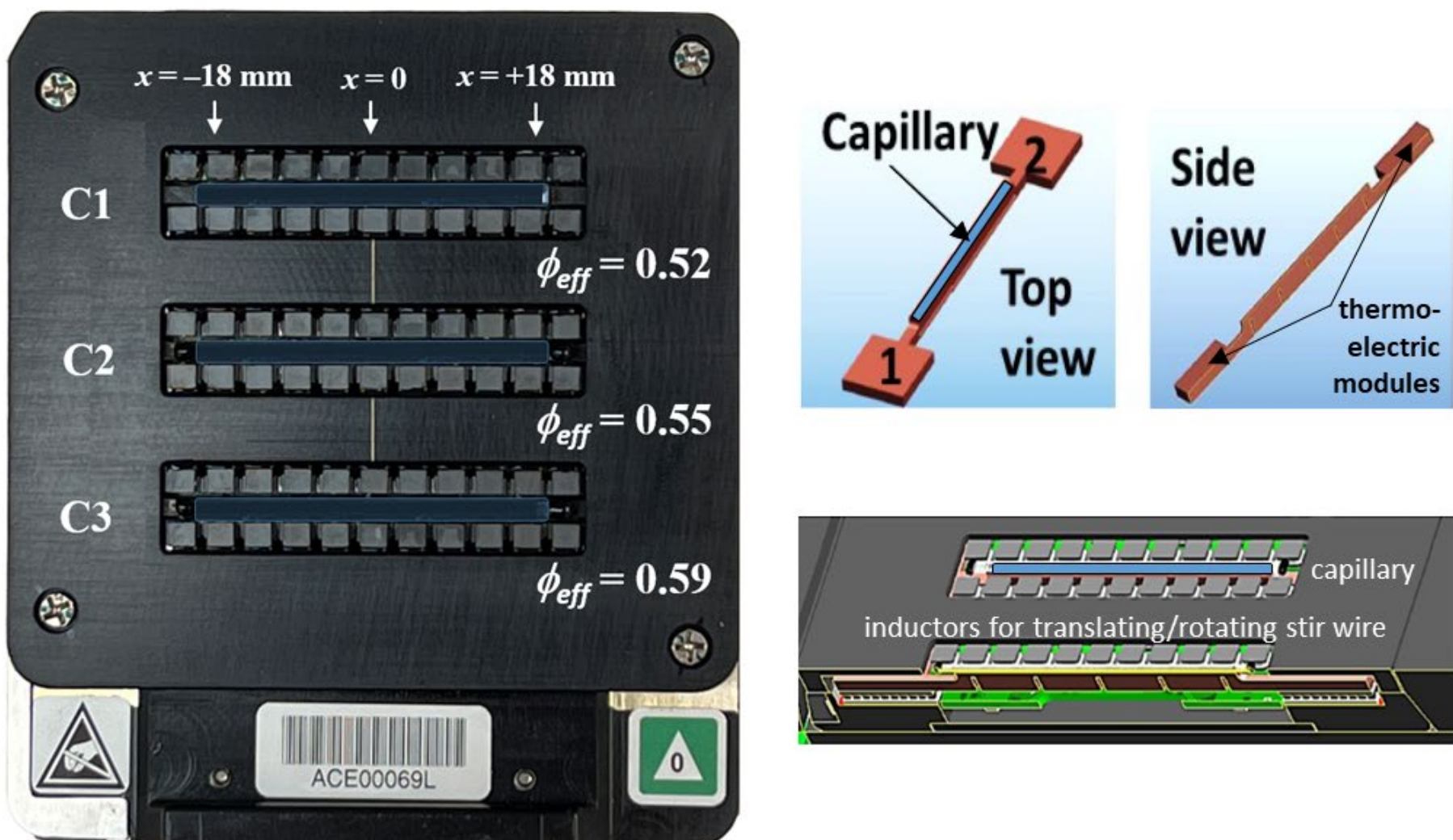

**Figure 3.** Sample module SN 2017. Capillary numbers are indicated along with the associated sample volume fractions (effective).

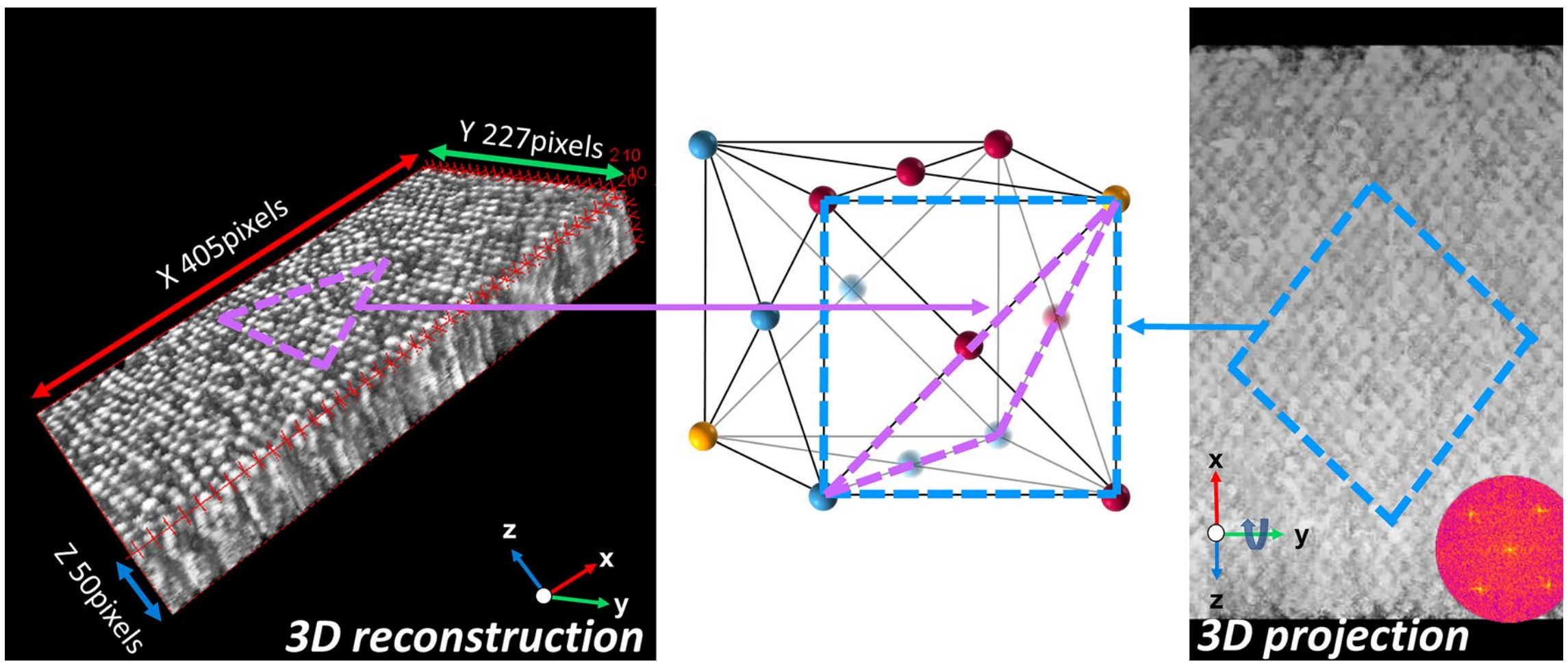

**Figure 4.** ACE-T11 LMM 3D confocal microscopy image (left) of large, defect-free colloidal photonic FCC crystal grown in C3. The 3D projection (right) shows the (100) face of a face-centered cubic crystal along with its 2-D Fourier transform.

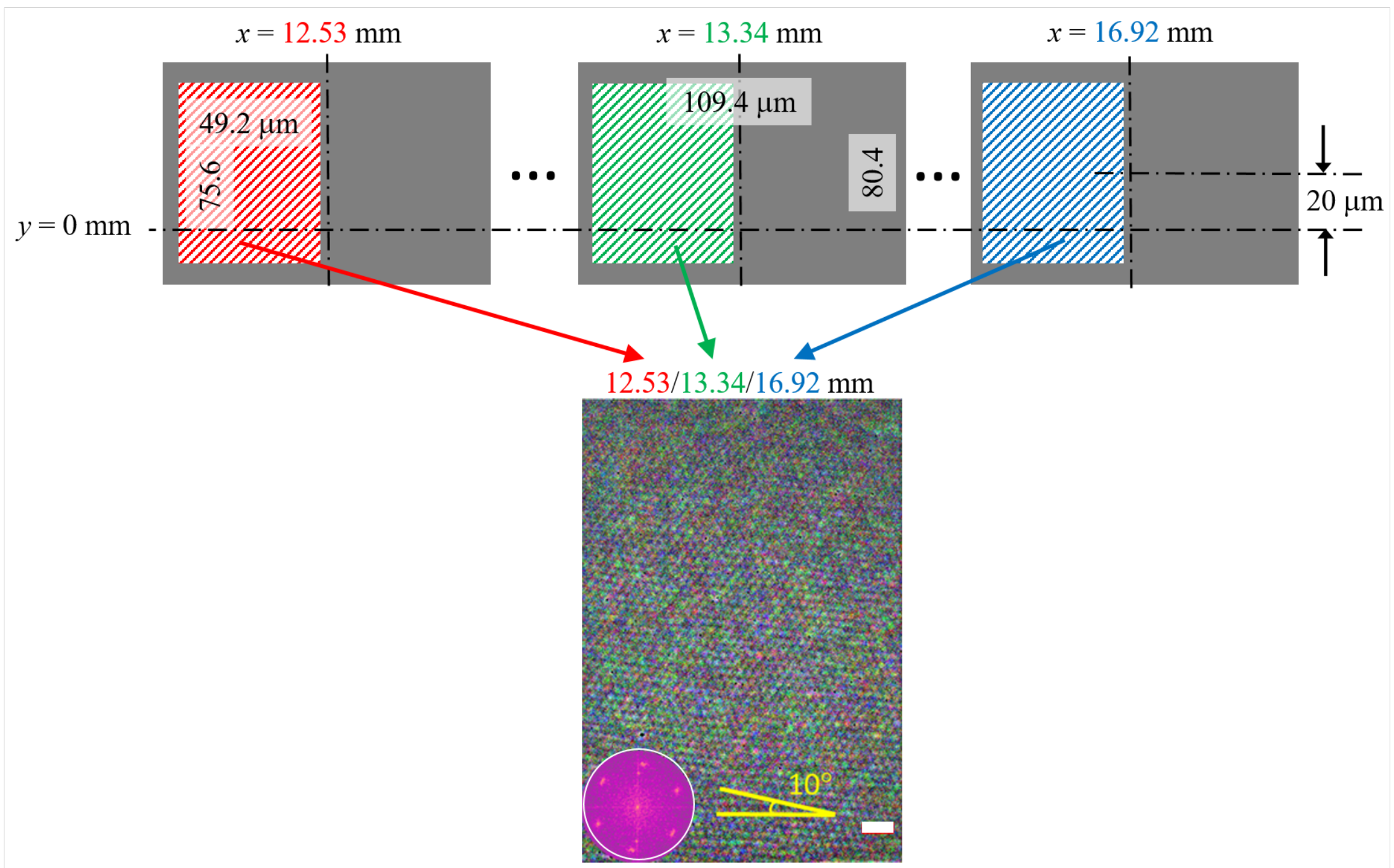

**Figure 5a.** Superposition of three *x-y* confocal microscopy images taken below the C3 capillary surface (*z* = 20 microns). The inset shows its 2-D Fourier transform, revealing the crystal structure and demonstrating structural coherence. The particle array is oriented approximately 10 degrees from the center line in capillary. Scale bar is 5 microns.

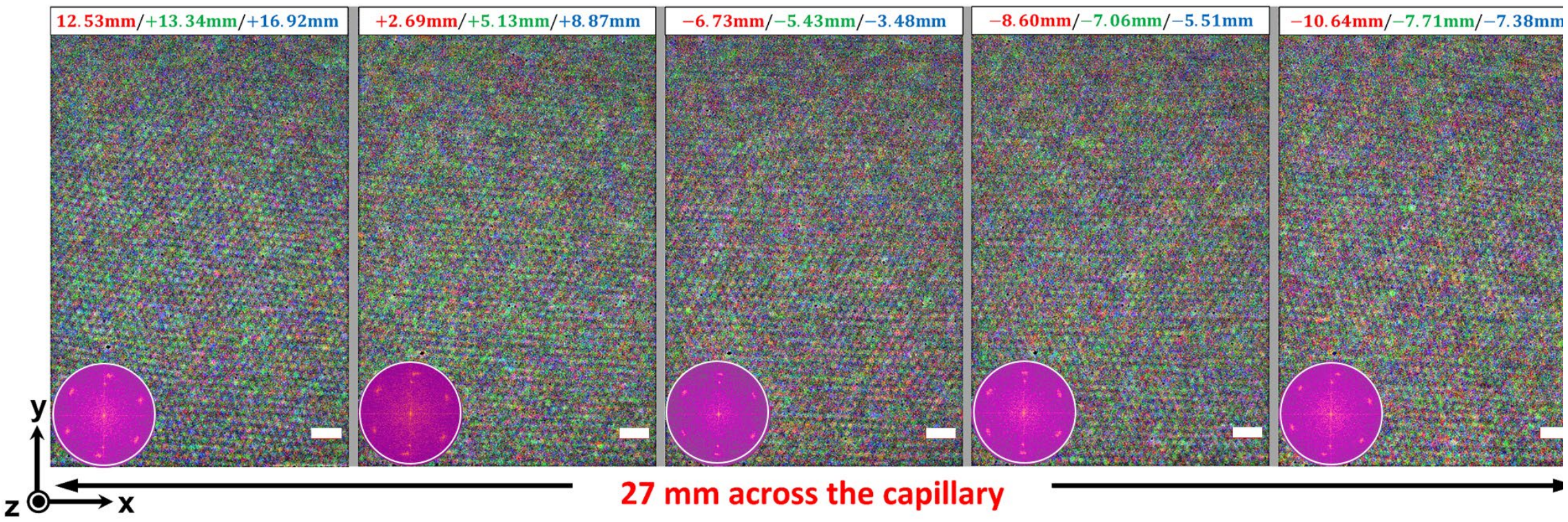

**Figure 5b.** Typical ACE-T11 LMM merged confocal images from an extended *x-y* slice used for constructing a confocal microscopy image of a large, defect free 3D colloidal photonic FCC crystal grown in space. The inset shows the corresponding 2-D Fourier transform, revealing the crystal structure. Scale bar is 5 microns.

## Acknowledgments

The research was supported by the NASA grant 80NSSC19K1655 to NJIT (B.K. and Q.L.) and the NSF grants 1832260 to NJIT (B.K. and Q.L.) and 1832291 to NYU (A.D.H. and P.M.C.) and NASA Contract 80GRC020D0003 for USRA at NASA GRC (W.V.M.). P.M.C. and A.D.H also acknowledge support from NASA grant NNX13AR67G. The Carl Zeiss FESEM instrument (NYU) was purchased with financial support from the MRI program of the National Science Foundation under Award DMR-0923251. The authors thank the ZIN Technologies Inc. team and the NASA GRC Payload Operations Center (GIPOC) for their support during flight operations. ImageJ is a Java-based image processing program developed at the National Institutes of Health and the Laboratory for Optical and Computational Instrumentation at University of Wisconsin.

## Author disclosure statement

No competing financial interests exist.